\documentclass[journal, a4paper, twoside]{IEEEtran}
\usepackage[margin = 1.41cm]{geometry}
\usepackage[utf8]{inputenc} 
\usepackage[T1]{fontenc}
\usepackage{url}
\usepackage{ifthen}
\usepackage{titlesec}
\usepackage{cite}
\usepackage[cmex10]{amsmath} 
\usepackage{amssymb}
\usepackage{graphicx}
\usepackage{epstopdf}
\usepackage{epsfig}
\usepackage{mathtools}
\DeclarePairedDelimiter\ceil{\lceil}{\rceil}

\usepackage{multirow}
\usepackage{array}
\usepackage{makecell}
\usepackage{tikz}
\usepackage{diagbox}
\usepackage{adjustbox}
\usepackage{caption, multirow, makecell}
\usepackage{algorithm}
\usepackage{algorithmic}
\usepackage{comment}
\usepackage[english]{babel}

\newtheorem{thm}{Theorem}
\newtheorem{lem}{Lemma}

\newtheorem{corollary}{Corollary}

\newtheorem{defn}{Definition}

\newtheorem{exmp}{Example}

\setcellgapes{3pt}

\newtheorem{rem}{Remark}

\begin{document}
	\title{Multi-Access Coded Caching with Demand Privacy} 
	\author{%
		\IEEEauthorblockN{K. K. Krishnan Namboodiri and B. Sundar Rajan \IEEEauthorrefmark{1}}\\
		\IEEEauthorblockA{\IEEEauthorrefmark{1}Department of Electrical Communication Engineering, Indian Institute of Science, Bengaluru\\
			E-mail: \{krishnank, bsrajan\}@iisc.ac.in}
	}
	\maketitle
	%%%%%%%%%%%%	
\begin{abstract}
The demand private coded caching problem in a multi-access network with $K$ users and $K$ caches, where each user has access to $L$ neighbouring caches in a cyclic wrap-around manner, is studied. The additional constraint imposed is that one user should not get any information regarding the demands of the remaining users. A lifting construction of demand private multi-access coded caching scheme from conventional, non-private multi-access scheme is introduced. The demand-privacy for a user is ensured by placing some additional \textit{keys} in a set of caches called the \textit{private set} of that user. For a given $K$ and $L$, a technique is also devised to find the private sets of the users.
\end{abstract}
\begin{IEEEkeywords}
Coded Caching, Multi-Access Network, Demand Privacy.
\end{IEEEkeywords}

	\IEEEpeerreviewmaketitle
	\section{Introduction}
	\label{intrn}
	
	Broadband data consumption has witnessed revolutionary growth over the past few years. The growing number of users and their increasing appetite for high data rate content leads to network traffic congestion, particularly during peak traffic periods. At the same time, the resources are often underutilized during off-peak periods. The idea of coded caching came to light after the seminal work of Maddah Ali and Niesen \cite{MaN}, in which it was shown that by utilizing multicast opportunities, coded caching could achieve significant gain over the conventional uncoded caching. The proposed set-up consists of a central server, having a library of $N$ files, connected to $K$ users through an error-free broadcast link. Each user is equipped with a dedicated cache, which can store $M$ out of the $N$ files in the placement phase. Each user reveals their demand during the delivery phase. Then, the server broadcasts coded symbols to all the users over the shared link. The objective is to design the placement and the delivery phases jointly such that the load of the shared link in the delivery phase is minimized. 
		
	However, in practical scenarios such as in cellular networks, the users can have access to multiple caches when their coverage areas overlap. Incorporating this possibility, coded caching problem has been extended to multi-access set-up recently \cite{SPE,ReK,SaR,MaR,ReK2,CLWZC,SaR2}, with $K$ users and $K$ caches, where each user can access $L$ neighbouring caches in a cyclic wrap-around fashion. 
	\begin{comment}
	In \cite{SPE}, the authors have proposed an optimal multi-access coded caching scheme for $L = \frac{K-1}{KM/N}$. A colouring based delivery scheme is proposed in \cite{ReK} and achieves a rate $K(1-\frac{LM}{N})^2$ for a cache memory $M$. In \cite{SaR,MaR} and \cite{SaR2}, the authors construct multi-access coded caching schemes with linear subpacketization. The construction of multi-access schemes from PDAs is shown in \cite{CLWZC} and \cite{SaR2}. The authors establish a connection between index coding and multi-access coded caching in \cite{ReK2}, and give a tighter upper bound on the optimal rate-memory trade-off. 
	\end{comment}
	
	One of the drawbacks of the aforementioned coded caching schemes is that the demand of every user will be revealed to the other users connected to the server in the delivery phase. The issue of demand-privacy has been studied in the dedicated cache networks in \cite{WaG,Kam,AST,GRKDK,NaR,YaT}. %Privacy for the user demands is ensured by the addition of virtual users in \cite{Kam,AST,GRKDK,NaR}. In \cite{YaT}, a demand-private coded caching scheme is constructed from linear function retrieval problem and brought down the subpacketization number drastically. 
	Recently, the multi-access coded caching problem was studied incorporating demand privacy constraint in \cite{LWCC}. In the multi-access coded caching scheme in \cite{LWCC}, the server stores some additional keys in the caches to ensure privacy for the user demands. The general scheme given in Theorem 1 in \cite{LWCC} ensures demand privacy when $L\leq \ceil{\frac{K}{2}}$, and fails to satisfy the demand-privacy condition when $L> \ceil{\frac{K}{2}}$. 
	 
	In this work, we focus on multi-access coded caching schemes ensuring privacy for the user demands. The main contributions of this paper can be summarized as follows,
	\begin{itemize} 
	\item A sufficient condition for the existence of multi-access coded caching schemes with demand privacy is presented. Using the idea of the private sets of the users, a lifting construction of the multi-access coded caching schemes ensuring demand privacy is introduced  (Section \ref{MainResults}, Theorem \ref{Lifting}).
	\item The notion of the private set of a user in a multi-access network is introduced. An algorithm to find private sets of the users is presented, and thereby, an upper bound on the size of the smallest private set is derived (Section \ref{MainResults}, Lemma \ref{sizeprivateset}).
	\end{itemize}  
	\subsection{Notations}
For a positive integer $n$, $[n]$ denotes the set $ \left\{1,2,\hdots,n\right\}$.
For any two integers, $i$ and $K$, 
$$
<i>_K =
\begin{cases}
	i\text{ }(mod\text{ }K) & \text{if $i$ $(mod$ $K) \neq0$. }\\
	K & \text{if $i$ $(mod$ $K) =0$.}
\end{cases}      
$$
For integers $a,b$ and $K$ such that $a,b\leq K$,
$$
[a:b]_K =
\begin{cases}
	\{a,a+1,\hdots,b\} & \text{if $a \leq b$ }\\
	\{a,a+1,\hdots,K,1,2\hdots,b\} & \text{if $a>b$ }
\end{cases} 
$$
For a set $\mathcal{S}$, $|\mathcal{S}|$ denotes the cardinality of set $\mathcal{S}$. The notation '$\oplus$' represents the Exclusive OR (XOR) operation between two bits. It denotes the element-wise XOR operation if the operands are binary vectors. 
	
	\section{System Model}
\begin{figure}[t]
		\begin{center}
			\captionsetup{justification = centering}
			\captionsetup{font=small}
			\includegraphics[width = 0.8\columnwidth]{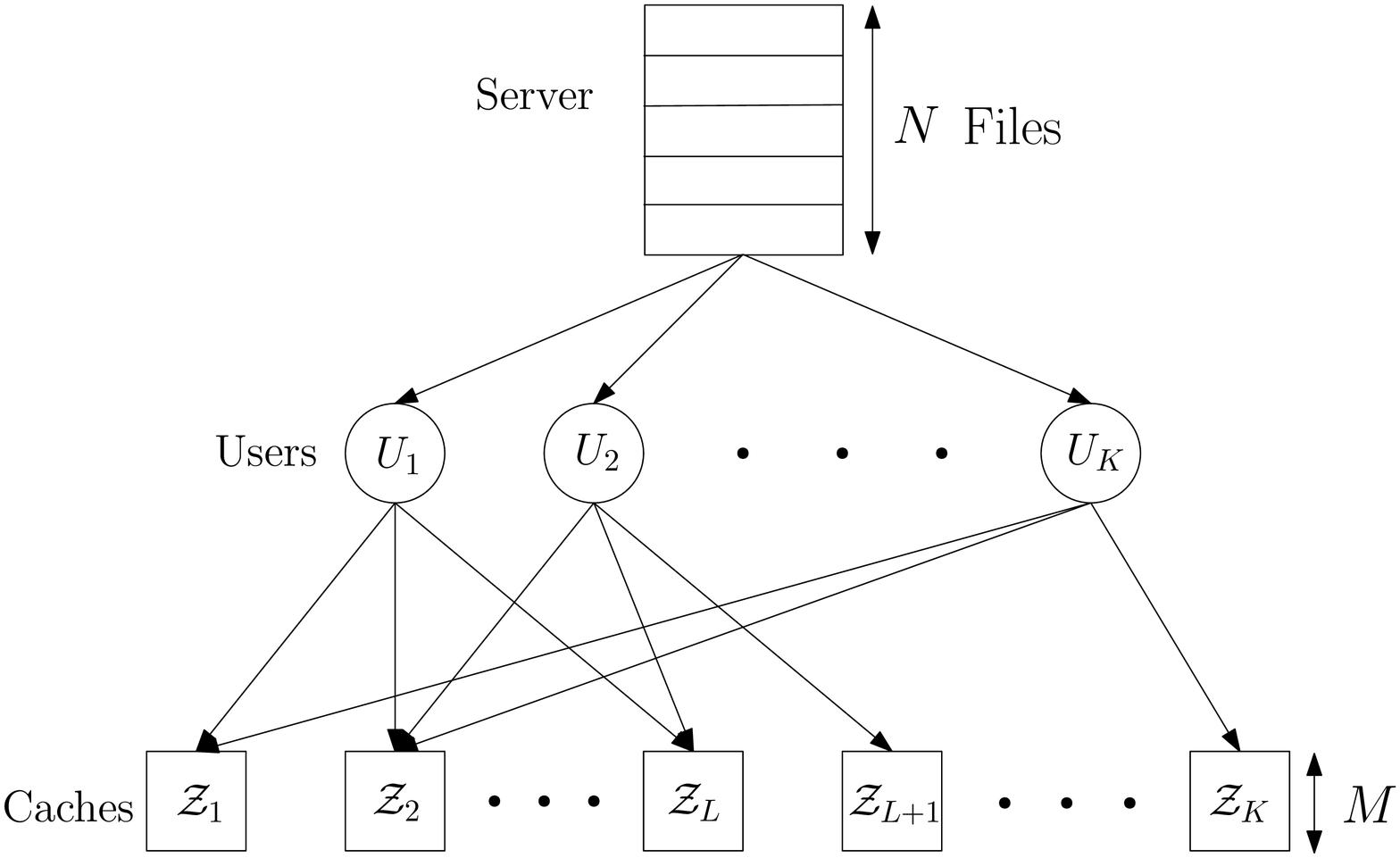}
			\caption{$(K,L,N)$ Multi-access Coded Caching Network. }
			\label{MACCNetwork}
		\end{center}
	\end{figure}

	The system model consists of a central server with a library of $N$ independent files, $\mathbf{W}=\left\{W_1,W_2,\hdots,W_N\right\}$ each of size $F$ bits, connected to $K$ users $\left\{U_1,U_2,\hdots,U_K\right\}$  through an error-free broadcast link. There are $K$ caches $\left\{\mathcal{Z}_1,\mathcal{Z}_2,\hdots,\mathcal{Z}_K\right\}$ in the system, each of capacity $MF$ bits, where $0\leq M \leq N$. Each user has access to $L<K$ caches in a cyclic wrap-around manner. $\mathcal{L}_k$ denotes the caches accessible for user $U_k$, and $\mathcal{L}_k = \left\{\mathcal{Z}_k,\mathcal{Z}_{<k+1>_K},\hdots,\mathcal{Z}_{<k+L-1>_K}\right\}$. Note that, $|\mathcal{L}_k|=L, \text{ }\forall k\in [K]$. A system under the aforementioned setting is called a $(K,L,N)$ multi-access network (Figure \ref{MACCNetwork}).	The coded caching schemes under this model have been discussed in \cite{SPE,ReK,SaR,MaR,ReK2,CLWZC,SaR2}. 	When $L=1$, this setting reduces to the problem setting in \cite{MaN}.
	
	A $(K,L,N)$ multi-access coded caching scheme works in two phases. In the \textit{placement phase}, the server fills the caches with the file contents without knowing the user demands. $Z_k$ denotes the content stored in the cache $\mathcal{Z}_k$. In the \textit{delivery phase}, each user requests a single file from the server through a private link between the user to the server without revealing to the peer users. Let $\mathcal{D}_k$ be the random variable denoting the demand of user $U_k$, and $\mathcal{D} = \left\{\mathcal{D}_1,\mathcal{D}_2,\hdots,\mathcal{D}_K\right\}$ are all uniformly and independently distributed over the set $[N]$. Let $\mathcal{D}_{\tilde{k}}$ denote all the demands but $\mathcal{D}_k$. That is, $\mathcal{D}_{\tilde{k}} = \mathcal{D}\backslash \{\mathcal{D}_k\}$. After knowing the demands of all the users, the server makes a transmission $X$ of size $RF$ bits. The non-negative real number $R$ is said to be the rate of the transmission. Each user $U_k$, $k\in [K]$ should be able to decode the demanded file from the transmission $X$ and the contents in the caches in $\mathcal{L}_k$. That is,
	\begin{equation}
		\label{Correctness}
		\text{[Decodability]     } H(W_{\mathcal{D}_k}|Z_{\mathcal{L}_k},\mathcal{D}_k,X)=0,\text{  } \forall k\in [K],
	\end{equation}
		where, $Z_{\mathcal{L}_k}$ denotes the contents stored in the caches in $\mathcal{L}_k$.
	\begin{defn}
		For a $(K,L,N)$ multi-access coded caching scheme, a memory-rate pair $(M,R)$ is said to be achievable if the scheme for cache memory $M$ satisfies the decodability condition in \eqref{Correctness} with a rate less than or equal to $R$ for every possible realization of $\mathcal{D}$.
		
		 The optimal rate-memory trade-off is defined as, 
		\begin{equation}
			\label{OptimalTradeoff}
			R^*(M) = inf \left\{R: (M,R) \text{ is achievable}\right\}.
		\end{equation}

	\end{defn}

	In addition to the decodability condition in \eqref{Correctness}, the demand-privacy condition is imposed on the $(K,L,N)$ multi-access coded caching scheme, as follows,
	\begin{equation}
		\label{DemandPrivacy}
		\text{[Demand privacy]       } I(\mathcal{D}_{\tilde{k}};Z_{\mathcal{L}_k},\mathcal{D}_k,X|\mathbf{W}  )=0, \text{  } \forall k\in [K].
	\end{equation} 
	We assume that the random variables $\mathcal{D}$ and $\mathbf{W}$ are all independent of each other. Thus, we can rewrite the demand privacy condition in \eqref{DemandPrivacy} as,
	\begin{equation}
		\label{DemandPrivacy_Rewritten}
		I(\mathcal{D}_{\tilde{k}};Z_{\mathcal{L}_k},\mathcal{D}_k,X,\mathbf{W}  )=0, \text{  } \forall k\in [K].
	\end{equation} 
	That is, user $U_k$ should be completely uncertain about the demanded file indices of the remaining users, given all the information $U_k$ has. The condition should hold even if $U_k$ has access to all the files $\mathbf{W}$. Note that, we are not considering the possibility of user collusion in the demand privacy condition.
	\begin{defn}
		For a $(K,L,N)$ multi-access coded caching scheme, a memory-rate pair $(M,R)$ is said to be achievable with demand privacy, if the scheme for cache memory $M$ satisfies the decodability condition in \eqref{Correctness} and the demand-privacy condition in \eqref{DemandPrivacy} with a rate less than or equal to $R$ for every possible realization of $\mathcal{D}$. 
		\begin{comment}
		The optimal rate-memory trade-off with demand privacy is defined as, 
		\begin{equation}
			\label{OptimalTradeoffDP}
			R_{dp}^*(M) = inf \{R: (M,R) \text{ is achievable with demand privacy}\}.
		\end{equation}
		\end{comment}
	\end{defn}

	The multi-access coded caching schemes satisfying the demand-privacy condition in addition to the decodability condition are called demand-private schemes, and the schemes that do not satisfy the demand-privacy condition are called non-private schemes.  
	%For SUCICP(SC), the broadcast rate $\ell^* \leq U+D+1$ and the broadcast rate is optimal if $U$ or $D$ is zero.
	\section{Main Results}
	\label{MainResults}
	
	In this section, we construct the demand-private coded caching schemes for $(K,L,N)$ multi-access networks. %Theorem \ref{Baseline} (Baseline Theorem) gives a $(K,L,N)$ multi-access coded caching scheme with demand privacy. 
	\begin{thm}(Baseline Theorem)
		\label{Baseline}
		For a $(K,L,N)$ multi-access coded caching scheme, for $0\leq M\leq \frac{N}{L}$, the rate $R(M)=N-LM$ is achievable with demand privacy.
		
	\end{thm}
	\begin{IEEEproof}
		We consider the memory regime $0\leq M\leq \frac{N}{L}$. In the placement phase, the server divides the file $W_n$, $n\in [N]$ into two non-overlapping subfiles $W_n^{(1)}$ and $W_n^{(2)}$ with $|W_n^{(1)}|=\frac{LMF}{N}$ bits and $|W_n^{(2)}|=(1-\frac{LM}{N})F$ bits. The subfile $W_n^{(1)}$ is further divided into $L$ non-overlapping subfiles of equal size, $\frac{MF}{N}$ bits. So, $W_n^{(1)} = \{W_{n,1}^{(1)},W_{n,2}^{(1)},\hdots,W_{n,L}^{(1)}\}$ for all $n\in [N]$. Now, consider an adjacent independent row (AIR) matrix $A_{K\times L}$. An AIR matrix of size $m\times n$ ($m\geq n$) is a zero-one matrix with the property that any $n$ adjacent rows (with cyclic wrap-around) of $A$ are linearly independent over $\mathbb{F}_q$ (irrespective of the field size $q$)\cite{VaR}. Using the AIR matrix $A_{K\times L}$, encode the subfiles $[W_{n,1}^{(1)},W_{n,2}^{(1)},\hdots,W_{n,L}^{(1)}]$. That is,
		\begin{equation*}
			\begin{aligned}
				\mathbf{C} &:= [\mathcal{C}_{n,1},\mathcal{C}_{n,2},\hdots,\mathcal{C}_{n,K}]\\
				&= [W_{n,1}^{(1)},W_{n,2}^{(1)},\hdots,W_{n,L}^{(1)}] A^T. 
			\end{aligned}
		\end{equation*} 
		The server fills the cache $\mathcal{Z}_k,k\in [K]$ with the coded subfile $\mathcal{C}_{n,k}$ of all the files $n\in [N]$. The size of a coded subfile is the same as the size of a subfile of $W_n^{(1)}$. Which means, $|C_{n,k}| = \frac{MF}{N}$ for all $k\in [K]$ and $n\in[N]$. So, the memory constraint of all the caches is met by the given placement strategy.
		 
		 In the delivery phase, irrespective of the demand vector, the server broadcasts $W_{n}^{(2)}$ for all $n\in [N]$. Therefore, the rate of the transmission is $N-LM$. 
		
		The claim is that each user can decode the demanded file using the transmission and the accessible cache contents. User $U_k$ directly receives $W_{d_k}^{(2)}$ from the broadcast transmission. User $U_k$ has access to $L$ adjacent caches (with a possible cyclic wrap-around), and $U_k$ possesses the coded subfiles $\mathcal{C}_{n,j}$ for $j\in [k:<k+L-1>_K]_K$ and for all $n\in [N]$. Since, every consecutive  $L$ columns of $A^T$ are linearly independent, user $U_k$ can get all the $L$ subfiles of $W_{n}^{(1)}$ for all $n\in [N]$. In short, $U_k$ can decode $W_{n}^{(1)}$ for every $n\in [N]$. That is, all the users can decode all the files from the transmissions and the cache contents. In addition to that, the placement and delivery are independent of the demand vector. Hence the demand privacy for the users is also guaranteed.
		
		This completes the proof of Theorem \ref{Baseline}.    
	\end{IEEEproof}

 Now, we present a lifting construction of the multi-access coded caching with demand privacy. Before dealing with the general construction, we give an example for the multi-access scheme with $K=3,L=2$ and $\frac{M}{N}=\frac{1}{3}$. 
	
	\begin{exmp}
		\label{example}
		Consider a $(K=3,L=2,\frac{M}{N}=\frac{1}{3})$ multi-access coded caching scheme. We assume that $N$ files ($N\geq 3$) are available with the server. First we will illustrate the case without demand privacy. In the placement phase, the server divides each file into $3$ equal sized subfiles with size $F/3$ bits. That is, $W_n = \{W_{n,1},W_{n,2},W_{n,3}\}$ for all $n\in [N]$. Cache $\mathcal{Z}_j$, $j\in \{1,2,3\}$ is filled with $W_{n,j}$ for all $n\in [N]$. By the placement, user $U_1$ gets the subfiles $W_{n,1}$ and $W_{n,2}$ of all the files. Similarly, user $U_2$ has $W_{n,2}, W_{n,3}$ and $U_3$ has $W_{n,3}$ and $W_{n,1}$ for all $n\in [N]$. Let $[d_1,d_2,d_3]$ be the demand vector. Then the server transmits the coded subfile, $W_{d_1,3}\oplus W_{d_2,1} \oplus W_{d_3,2}$. 
		
		Now, we construct a demand private  multi-access coded caching scheme with $K=3,L=2$ and $M = \frac{N+2}{3}$. In the placement phase, the server generates vectors $\mathbf{p}_k^{(\alpha)}$ for every $k\in \{1,2,3\}$ and $\alpha \in \{1,2\}$, which are all uniformly and independently drawn from the set of binary vectors of length $N$. Let $\mathbf{p}_k^{(\alpha)}:= [p_{k,1}^{(\alpha)},p_{k,2}^{(\alpha)},\hdots,p_{k,N}^{(\alpha)}]$. Then the server creates 6 coded subfiles as follows,
		
		$$S_1^{(\alpha)} = \bigoplus_{n\in [N]} p_{1,n}^{(\alpha)}W_{n,3}, \hspace{0.5cm} \alpha \in \{1,2\}.$$
		$$S_2^{(\alpha)} = \bigoplus_{n\in [N]} p_{2,n}^{(\alpha)}W_{n,1}, \hspace{0.5cm} \alpha \in \{1,2\}.$$
		$$S_3^{(\alpha)} = \bigoplus_{n\in [N]} p_{3,n}^{(\alpha)}W_{n,2}, \hspace{0.5cm} \alpha \in \{1,2\}.$$
		 $S_1^{(1)}$ and $S_1^{(2)}$ are linear combinations of the subfiles $W_{n,3}, \forall n\in [N]$, the subfiles that user $U_1$ has no access to in the uncoded form. Similarly, $U_2$ does not have access to the subfiles $W_{n,2}$ in the uncoded form, and $S_2^{(1)}$ and $S_2^{(2)}$  are precisely the linear combinations of those subfiles. The same is the case with user $U_3$. The cache placement in the demand-private scheme is as follows,
		$$Z_1 = \{W_{n,1},\forall n\in [N], S_1^{(1)}, S_3^{(2)} \},$$
		$$Z_2 = \{W_{n,2},\forall n\in [N], S_2^{(1)}, S_1^{(2)} \},$$
		$$Z_3 = \{W_{n,3},\forall n\in [N], S_3^{(1)}, S_2^{(2)} \}.$$
	%	\noindent Figure \ref{imageK=3} illustrates the contents stored in each of the three caches.\\
		
		 We define a key for a user as follows,
		\begin{equation}
			\label{KeyforUser}
			\mathbb{K}_k:= S_k^{(1)}\oplus S_k^{(2)}, \hspace{1cm} k\in \{1,2,3\}.
		\end{equation} 
		%\begin{figure}[h]
		%	\begin{center}
		%		\captionsetup{justification = centering}
		%		\includegraphics[width = 0.8 \columnwidth]{K=3dp.eps}
		%		\caption{Contents stored in each of the three caches.}
		%		\label{imageK=3}
		%	\end{center}
		%\end{figure}
		Notice that only $U_1$ can calculate $\mathbb{K}_1$ since no other user has access to both $\mathcal{Z}_1$ and $\mathcal{Z}_2$. Similarly, only $U_2$ can calculate $\mathbb{K}_2$ and only $U_3$ can calculate $\mathbb{K}_3$. Note that the possibility of the user collusion is not considered here. 
		
		Let the demand vector be $[d_1,d_2,d_3]$. In the delivery phase, the server makes the transmission,
		$$X = (Q,\tilde{X}),$$
		$$\text{where, }\hspace{0.3cm}Q = [\mathbf{q}_1, \mathbf{q}_2, \mathbf{q}_3],$$
		$$\text{and }\hspace{0.1cm}\tilde{X} = \left(W_{d_1,3}\oplus \mathbb{K}_1\right)\oplus \left( W_{d_2,1} \oplus \mathbb{K}_2 \right) \oplus \left(  W_{d_3,2} \oplus  \mathbb{K}_3\right),$$
	    where, $\mathbf{q}_k = \mathbf{p}_k^{(1)} \oplus \mathbf{p}_k^{(2)}\oplus \mathbf{e}_{d_k}$ for $k \in \{1,2,3\}$, and $\mathbf{e}_n\in \mathbb{F}_2^N$ is the standard unit vector with a one in the $n$-th position and zero elsewhere. Notice that,
		$$W_{d_1,3}\oplus \mathbb{K}_1  = \bigoplus_{n\in [N]} q_{1,n} W_{n,3},$$
		$$W_{d_2,1}\oplus \mathbb{K}_2 = \bigoplus_{n\in [N]} q_{2,n} W_{n,1},$$
		$$W_{d_3,2}\oplus \mathbb{K}_3 = \bigoplus_{n\in [N]} q_{3,n} W_{n,2}.$$
		User $U_1$ can compute $W_{d_2,1}\oplus \mathbb{K}_2$ and $W_{d_3,2}\oplus \mathbb{K}_3$ by using the subfiles $W_{n,1},W_{n,2}, n\in [N]$, and the second and the third columns of $Q$. Also, $U_1$ can calculate $\mathbb{K}_1$. So $U_1$ can decode $W_{d_1,3}$. Similarly, the users $U_2$ and $U_3$ can also decode the subfiles they need. The demand privacy is guaranteed since a user does not know the key of the remaining two users, and all the three columns of $Q$ are uniformly and independently distributed over $\mathbb{F}_2^N$. $U_2$ knows $S_1^{(2)}$ and can find out $\mathbf{p}_1^{(2)}$, but it does not know $\mathbf{p}_1^{(1)}$. Similarly, $U_3$ can find $\mathbf{p}_1^{(1)}$ from $S_1^{(1)}$ but it does not know $\mathbf{p}_1^{(2)}$. So, in the second and third users' point of view, $d_1\sim unif{[N]}$. But notice that, if $U_2$ and $U_3$ collude, they can know $d_1$ without any ambiguity. But, in the demand privacy condition \eqref{DemandPrivacy}, we are not incorporating the possibility of user collusion.
		
		The server stores $N+2$ subfiles in all the caches, each of size $F/3$ bits. Therefore, the cache memory $M$ is $\frac{N+2}{3}$. Note that, irrespective of $N$, the number of additional subfiles stored in a cache is two, compared to the non-private scheme. Therefore, the file-normalized memory $\frac{M}{N}$ reduces and tends to $\frac{1}{3}$ as the number of files increases. Or in other words, the fraction of the additional memory used for ensuring demand-privacy is very less when $N\gg L$. 
		
		In the delivery phase, the main payload is $\tilde{X}$  which is of the size $F/3$ bits. The matrix $Q$ can be transmitted using $3N$ bits, that does not scale with $F$. Thus, the scheme achieves the memory-rate pair $(\frac{N+2}{3},\frac{1}{3})$.
		\end{exmp}
	\subsection{Lifting construction of Demand private schemes}
	To ensure privacy for the user demands, it is required to store some keys in the caches. The keys for user $U_k$ are the linear combinations of the subfiles that are not placed in the uncoded form in the caches in $\mathcal{L}_k$. In Example \ref{example}, the key for user $U_1$ is $\mathbb{K}_1$, which is the sum of $S_1^{(1)}$ and $S_1^{(2)}$. The only user who can access both $S_1^{(1)}$ and $S_1^{(2)}$ is $U_1$ since no other user has access to both $\mathcal{Z}_1$ and $\mathcal{Z}_2$. In general, the keys for user $U_k$ should remain private between $U_k$ and the server. So, one naive approach is to place the keys in a distributed manner in all the $L$ caches in $\mathcal{L}_k$, since no other user has access to all of those caches. But, for most of the values of $K$ and $L$, we can do significantly better than that. This means that, during the placement phase, the keys for $U_k$ will be stored only in a subset of the set of caches $\mathcal{L}_k$.  
	\begin{defn}[Private Set]
		For user $U_k$, $k\in [K]$, a private set $\mathcal{L}_k^p$ is a subset of $\mathcal{L}_k$ such that $\mathcal{L}_k^p \nsubseteq \mathcal{L}_{k'}$ for every $k'\neq k$.
	\end{defn}  

    For example, if $K=4$ and $L=3$, the private set of $U_1$ is $\mathcal{L}_1^p = \{\mathcal{Z}_1,\mathcal{Z}_2,\mathcal{Z}_3\}=\mathcal{L}_1$. In this case, it can be verified that $\mathcal{L}_k^p = \mathcal{L}_k$ for all $k\in [4]$. Now, consider the case $K=5$ and $L=3$. Note that for user $U_1$, $\mathcal{L}_1^p =\{\mathcal{Z}_1,\mathcal{Z}_3\}\subset \mathcal{L}_1$ is a private set of $U_1$, since $\mathcal{L}_1^p \nsubseteq \mathcal{L}_k$ for any $k \in \{2,3,4,5\}$.  
    
    For any $K$ and $L$, $\mathcal{L}_k$ is a private set of user $U_k$ with cardinality $L$. In general, private set of a user is not unique. If $\mathcal{L}_1^p = \{\mathcal{Z}_{\ell_1},\mathcal{Z}_{\ell_2},\hdots,\mathcal{Z}_{\ell_t}\}$ is a private set of $U_1$ with cardinality $t$ such that $\ell_1<\ell_2<\hdots\ell_t\leq L$, then $\mathcal{L}_k^p = \{\mathcal{Z}_{<\ell_1+k-1>_K},\mathcal{Z}_{<\ell_2+k-1>_K},\hdots,\mathcal{Z}_{<\ell_t+k-1>_K}\}$ is a private set of $U_k$, because of the circular symmetry of the network model. Let $t^*$ be the size of the smallest private set of $U_1$. Then, it can be seen that the same $t^*$ is the size of the smallest private set of all the users. In other words, $t^*$ is fixed for a given $K$ and $L$. 

    In further discussions, while referring to non-private multi-access schemes, we assume that the placement is uncoded and file symmetric. By file symmetric placement, we mean that, if a subfile $W_{n,j}$ of a file $W_n$ is kept in cache $\mathcal{Z}_k$, then the subfiles $W_{n,j}$ for all $n\in[N]$ is kept in $\mathcal{Z}_k$. 
    
    In the following theorem, we give a sufficient condition for the existence of a $(K,L,N)$ demand-private multi-access coded caching scheme.
	\begin{thm}
		\label{Lifting}
		If a memory-rate pair $(M,R)$ is achievable for a $(K,L,N)$ non-private multi-access coded caching scheme satisfying the condition C1, then the memory rate pair $(\tilde{M},\tilde{R}):= \left(M+t(1-\frac{LM}{N}),R\right)$ is achievable for $(K,L,N)$ demand private multi-access coded caching scheme, where $t$ is the size of the private set.
		\begin{comment}
		For a $(K,L,N)$ multi-access coded caching scheme, if a memory-rate pair $(M,R)$ is achievable satisfying the conditions C1 and C2, then the memory rate pair $(\tilde{M},\tilde{R}):= \left(M+t(1-\frac{LM}{N}),R\right)$ is achievable with demand privacy, where $t$ is the size of the private set.
		\end{comment}
		\begin{itemize}
			\begin{comment}
			\item C1: The placement is uncoded and file symmetric. By file symmetric placement, we mean that, if a subfile $W_{n,j}$ of the file $W_n$ is kept in cache $\mathcal{Z}_k$, then the subfiles $W_{n,j}$ for all $n\in[N]$ is kept in $\mathcal{Z}_k$.
			\end{comment}
			\item C1: The effective memory accessible for every user is $LM$. That is, for a user $U_k$, $k\in[K]$, $Z_j\cap Z_{j'}=\{\}$ for all $\mathcal{Z}_j,\mathcal{Z}_j' \in \mathcal{L}_k$ and $j\neq j'$. 
		\end{itemize}
	\end{thm}
	\begin{IEEEproof}	
		Assume that we have a $(K,L,N)$ non-private multi-access coded caching scheme that satisfies the condition C1, and achieves the rate $R$ at cache memory $M$. Let $\mathcal{L}_k^p = \{\mathcal{Z}_{<\ell_1+k-1>_K},\mathcal{Z}_{<\ell_2+k-1>_K},\hdots,\mathcal{Z}_{<\ell_t+k-1>_K}\}$ be a private set of $U_k,k\in[K]$ of size $t$.  Now, we construct a $(K,L,N)$ multi-access coded caching scheme with demand privacy, and achieves the same rate $R$ at cache memory $\tilde{M}=M+t(1-\frac{LM}{N})$.
		
		 \textit{Placement Phase:} The subpacketization of the files will remain the same as that of the non-private scheme. In the demand-private scheme, the cache placement is divided into two rounds.
		\begin{itemize}
			\item Round 1: The server follows the same cache placement in the non-private scheme.
			\item Round 2: The server uniformly and independently generates $Kt$ vectors from $\mathbb{F}_2^N$, $\mathbf{p}_k^{(\alpha)},k\in [K],\alpha \in [t]$. Let $\mathbf{p}_k^{(\alpha)}:=\left[p_{k,1}^{(\alpha)},p_{k,2}^{(\alpha)},\hdots,p_{k,N}^{(\alpha)}\right]$. For user $U_k$, $k\in [K]$, if a subfile $W_{n,j}$ is not available in the caches in $\mathcal{L}_k$ from Round 1, then store $\underset{n\in [N]}{\bigoplus} p_{k,n}^{(\alpha)}W_{n,j}$ in $\mathcal{Z}_{<\ell_\alpha+k-1>_K}$ for all $\alpha \in [t]$.   
		\end{itemize}
		
		In Round 1, the server stores $M/N$ fraction of all the files in all the caches. By the condition C1, each user will be accessing $\frac{LM}{N}$ fraction of all the files. But, a cache will be in the private set of $t$ users. Therefore, all the caches need an additional memory of $t(1-\frac{LM}{N})$ in Round 2. Hence the total cache memory is $M+t(1-\frac{LM}{N})$.
		
		\textit{Delivery Phase:} The transmission $X$ has two components, $X = (Q,\tilde{X})$. The server broadcasts the matrix $Q = \left[\mathbf{q}_1,\mathbf{q}_2,\hdots,\mathbf{q}_K\right]_{K\times N}$ to all the users, where 
		\begin{equation*}
		\mathbf{q}_k = \mathbf{e}_{d_k} \bigoplus \left(\underset{\alpha \in [t]}{\bigoplus}\mathbf{p}_k^{(\alpha)}\right), \hspace{0.3cm} \forall k\in [K].
	\end{equation*}
    
    Let $\mathbf{q}_k = [q_{k,1},q_{k,2},\hdots,q_{k,n}]$.
	Define,
	\begin{equation*}
		\widehat{W}_{d_k} \triangleq \bigoplus_{n\in[N]} q_{k,n}W_n, \hspace{0.3cm} \forall k\in [K].
	\end{equation*}	
		%$\widehat{W}_{d_k}$ is a linear combination of all the files, but the server treats $\widehat{W}_{d_k}$ as a new file. 
		
		The server treats the linear combination of the subfiles $\widehat{W}_{d_k}$ as a new file and makes the broadcast transmission according to the non-private scheme by assuming that user $U_k$ is requesting for file $\widehat{W}_{d_k}$ for all $k\in [K]$. This broadcast transmission is denoted as $\tilde{X}$.
		
		\textit{Decodability:} From the contents stored in Round 1 of the placement phase and from the transmission $\tilde{X}$, user $U_k$ can decode $\widehat{W}_{d_k}$ since the coefficients $q_{k,n}$ are available from the transmitted matrix $Q$. But user $U_k$ wants the file $W_{d_k}$. Some subfiles of all the files are available to $U_k$ from Round 1 of the placement phase. Now, user $U_k$, $k\in[K]$ can find every missing subfile $W_{d_k,j}$ as,
				\begin{equation*}
			W_{d_k,j} = \underbrace{\bigoplus_{n\in [N]}q_{k,n}W_{n,j}}_{\widehat{W}_{{d_k},j}}\bigoplus \left(\bigoplus_{\alpha \in [t]}\underbrace{\left(\bigoplus_{n\in [N]} p_{k,n}^{(\alpha)}W_{n,j}\right)}_{\text{Accessible for user $U_k$}}\right).
		\end{equation*}
	The coded subfile $\bigoplus_{n\in [N]} p_{k,n}^{(\alpha)}W_{n,j}$ is available for user $U_k$ in the cache $\mathcal{Z}_{<\ell_\alpha+k-1>_K}$.
	
	Now, the only thing left to prove is that the achievable scheme presented satisfies the demand-privacy condition. The keys to ensure the privacy for a user's demand is distributed across the private set of caches of that user. For user $U_k$, and for an index $j$ of a missing subfile, the key is defined as,
		\begin{equation}
			\mathbb{K}_k^{(j)} = \bigoplus_{\alpha \in [t]}\bigoplus_{n\in [N]} p_{k,n}^{(\alpha)}W_{n,j}.
		\end{equation}
		Notice that, from the point of view of user $U_{k'}$, $k'\neq k$, the $k^{th}$ column of the matrix $Q$ will be uniformly distributed over $\mathbb{F}_2^N$. Because, user $U_{k'}$ does not have access to all the caches in $\mathcal{L}_k^p$ and thus no other user can know the demand of $U_k$.
		
	 For user $U_k$, $ k\in [K]$ with a private set of caches $\mathcal{L}_k^p = \{\mathcal{Z}_{<\ell_1+k-1>_K},\mathcal{Z}_{<\ell_2+k-1>_K},\hdots,\mathcal{Z}_{<\ell_t+k-1>_K}\}$, define the set of vectors, $\mathcal{P}_k:=\left\{\mathbf{p}_{i}^{(\alpha)}|\mathcal{Z}_{<\ell_\alpha+i-1>_K}\in \mathcal{L}_k\right\} $. The vectors in $\mathcal{P}_k$ contain the coefficients of all the linear combinations of the subfiles stored in the caches in $\mathcal{L}_k$  (in Round 2 of the placement phase). 		Define the matrix, $R := \left[\mathbf{r}_1,\mathbf{r}_2,\hdots,\mathbf{r}_K\right]_{N\times K}$, where $\mathbf{r}_k = \underset{\alpha \in [t]}{\bigoplus} \mathbf{p}_k^{(\alpha)}$ for every $k\in [K]$. Note that, $\mathbf{q}_k = \mathbf{r}_k \oplus \mathbf{e}_{d_k}$. Also, $Q_{\tilde{k}}$ and $R_{\tilde{k}}$ denote the matrices $Q$ and $R$ with the $k^{th}$ column removed. Then,
		\begin{subequations}
			\begin{align}
				I(\mathcal{D}_{\tilde{k}};\mathcal{D}_k,X,Z_{\mathcal{L}_k},\mathbf{W})&= I(\mathcal{D}_{\tilde{k}};\mathcal{D}_k,Q,\tilde{X},Z_{\mathcal{L}_k},\mathbf{W}) \label{eqapa}\\
				&= I(\mathcal{D}_{\tilde{k}};\mathcal{D}_k,Q,Z_{\mathcal{L}_k},\mathbf{W})\label{eqapb}\\
				&\leq I(\mathcal{D}_{\tilde{k}};\mathcal{D}_k,Q,Z_{\mathcal{L}_k},\mathcal{P}_k,\mathbf{W})\\
				&= I(\mathcal{D}_{\tilde{k}};\mathcal{D}_k,Q,\mathcal{P}_k,\mathbf{W})\label{eqapd}\\
				&=I(\mathcal{D}_{\tilde{k}};\mathcal{D}_k,\mathcal{P}_k,\mathbf{W})\notag\\
				 &\qquad +I(\mathcal{D}_{\tilde{k}};Q|\mathcal{D}_k,\mathcal{P}_k,\mathbf{W})\\
				&=I(\mathcal{D}_{\tilde{k}};Q|\mathcal{D}_k,\mathcal{P}_k,\mathbf{W})\label{eqapf}\\
				&=I(\mathcal{D}_{\tilde{k}};Q_{\tilde{k}},\mathbf{q}_k|\mathcal{D}_k,\mathcal{P}_k,\mathbf{W})\\
				&=I(\mathcal{D}_{\tilde{k}};Q_{\tilde{k}}|\mathcal{D}_k,\mathcal{P}_k,\mathbf{W})\\
				&=H(Q_{\tilde{k}}|\mathcal{D}_k,\mathcal{P}_k,\mathbf{W})\notag\\
				 &\qquad-H(Q_{\tilde{k}}|\mathcal{D},\mathcal{P}_k,\mathbf{W})\\
				&= H(Q_{\tilde{k}}|\mathcal{P}_k)-H(R_{\tilde{k}}|\mathcal{D},\mathcal{P}_k,\mathbf{W})\\
				&= H(Q_{\tilde{k}}|\mathcal{P}_k)-H(R_{\tilde{k}}|\mathcal{P}_k)\\
				&=0.
			\end{align}
		\end{subequations} 
		 \eqref{eqapa} follows since $X = (Q,\tilde{X})$. $\tilde{X}$ is completely determined by matrix $Q$ and $\mathbf{W}$, \eqref{eqapb} holds. $Z_{\mathcal{L}_k}$ is a function of $\mathbf{W}$ and the vectors in $\mathcal{P}_k$. Therefore, \eqref{eqapd} follows. \eqref{eqapf} holds because $\mathcal{D}_{\tilde{k}}$ is independent of $\mathcal{D}_k$, $\mathcal{P}_k$ and $\mathbf{W}$. By the definition of private set, $\mathcal{P}_k$ contains $\mathbf{p}_k^{(\alpha)}$ for all $\alpha \in [t]$, but does not contain $\mathbf{p}_{k'}^{(\alpha)}$ for all $\alpha \in [t]$ for any $k'\neq k$. Hence $H(Q_{\tilde{k}}|\mathcal{P}_k)=H(R_{\tilde{k}}|\mathcal{P}_k)=N(K-1)$ bits. 
		
		This completes the proof of Theorem \ref{Lifting}.
	\end{IEEEproof}
\begin{rem}
	In the demand-private multi-access coded caching scheme given in \cite{LWCC}, irrespective of $K$ and $L$, the keys for user $U_k$ are always stored in the caches $\mathcal{Z}_k$ and $\mathcal{Z}_{<k+L-1>_K}$, and it is claimed that no user, except $U_k$, will have access to both these caches simultaneously. But, when $L> \ceil{\frac{K}{2}}$, the claim is not true. In that case, the user $U_{<k+L-1>_K}$ will also be accessing $\mathcal{Z}_k$ and $\mathcal{Z}_{<k+L-1>_K}$. For example, when $K=4$ and $L=3$, according to \cite{LWCC}, the keys for user $U_1$ will be stored in $\mathcal{Z}_1$ and $\mathcal{Z}_3$. At the same time, user $U_3$ is also accessing both $\mathcal{Z}_1$ and $\mathcal{Z}_3$. As a result, $U_3$ will get to know the demand of $U_1$.  
\end{rem}
\begin{rem}
	The non-private multi-access coded caching schemes in \cite{SPE,ReK,SaR,MaR,ReK2,CLWZC,SaR2} satisfy the condition C1. So, it is possible to derive demand-private schemes from those non-private schemes using the construction in the proof of Theorem \ref{Lifting}. 
\end{rem}

By identifying the private set with the smallest size, we can reduce the additional memory used for ensuring the demand privacy. In the following lemma, we give an upper bound on the size of the smallest private set by presenting a construction of the private sets of the users.
	\begin{lem}
		\label{sizeprivateset}
		For a given $K$ and $L$, the size of the smallest private set $t^*\leq \ceil{\frac{K-1}{K-L}}$.
	\end{lem}
	\begin{IEEEproof}
	We identify a set $\mathcal{L}_1^p \subseteq \mathcal{L}_1$ of size $\ceil{\frac{K-1}{K-L}}$ such that $\mathcal{L}_1^p\nsubseteq \mathcal{L}_k$ for all $k\neq 1$. From $\mathcal{L}_1^p$, we can find $\mathcal{L}^p_k$ for all $k\in [K]$. The construction of $\mathcal{L}^p_1$ is given in Algorithm \ref{alg1}. 
	
		\begin{algorithm}
			\caption{Private set construction for user $U_1$}
			\begin{algorithmic} 
				\label{alg1}
				\REQUIRE $K$, $L$
				\ENSURE $\mathcal{L}^p_1$
				\STATE \textbf{Initialize:} $i\gets 1$, $\mathcal{L}^p_1 \leftarrow \left\{\right\}$ 
				\WHILE{$i < L$}
				\STATE $\mathcal{L}^p_1 \leftarrow \mathcal{L}^p_1\cup \left\{\mathcal{Z}_i\right\}$
				\STATE $i\gets K-L+i$
				\ENDWHILE
				\STATE $\mathcal{L}^p_1 \leftarrow \mathcal{L}^p_1\cup \left\{\mathcal{Z}_L\right\}$
			\end{algorithmic}
		\end{algorithm}   
	First of all, $\mathcal{L}^p_1$ should contain $\mathcal{Z}_1$, since user $U_2$ has access to all the caches in $\mathcal{L}_1$ except $\mathcal{Z}_1$. 	
%Users $U_2$ to $U_{K-L+1}$ do not have access to $\mathcal{Z}_1$, but user $U_{k-L+2}$ has access to $\mathcal{Z}_1$. So, next we add $\mathcal{Z}_{K-L+1}$ to $\mathcal{L}^p_1$, since users from $U_{k-L+2}$ to $U_{2(K-L)+1}$ do not have access to that cache. 
Then, add the caches $\mathcal{Z}_{n(K-L)+1}$ to $\mathcal{L}^p_1$ for integer values of $n$ such that $n(K-L)+1<L$. Now, the set $\mathcal{L}^p_1 = \left\{\mathcal{Z}_{n(K-L)+1}\right\}_{n=0}^{\ceil{\frac{L-1}{K-L}}-1}$. Note that, $\ceil{\frac{L-1}{K-L}}(K-L)+1 \geq L$. The cache $\mathcal{Z}_{n(K-L)+1}$ is not accessible for the users $U_{n(K-L)+2}$ to $U_{(n+1)(K-L)+1}$. Therefore, $\mathcal{L}^p_1$ contains at least one cache that is not accessible for the users $U_2$ to $U_{\ceil{\frac{L-1}{K-L}}(K-L)+1}$. Finally, we add cache $\mathcal{Z}_L$ to $\mathcal{L}^p_1$, since the users $U_{L+1}$ to $U_K$ do not have access to $\mathcal{Z}_L$. By Algorithm \ref{alg1}, we make sure that there is at least one cache in $\mathcal{L}^p_1$ that is not accessible to the remaining users. Therefore, $\mathcal{L}^p_1$ is a private set of user $U_1$ with size $\ceil{\frac{K-1}{K-L}}$. For user $U_k$, $k\in [K]$ the private set is thus $\mathcal{L}^p_k =   \left\{\mathcal{Z}_{<n(K-L)+k>_K}\right\}_{n=0}^{\ceil{\frac{L-1}{K-L}}-1}\cup \left\{\mathcal{Z}_{<L+k-1>_K}\right\}$.

This completes the proof of Lemma \ref{sizeprivateset}. 	         
	\end{IEEEproof}
\begin{rem}
	\label{specialcase}
	For $2\leq L \leq \ceil{\frac{K}{2}}$, the size of the smallest private set is $t^{*}=2$, and $\mathcal{L}^p_k = \left\{\mathcal{Z}_k,\mathcal{Z}_{<k+L-1>_K}\right\}$. Similarly, when $L=K-1$, $t^{*}=K-1$ itself, ie; $\mathcal{L}^p_k = \mathcal{L}_k$ for all $k\in [K]$.
\end{rem}
	\begin{figure}[t]
	\begin{center}
		%	\captionsetup{justification = centering}
		\captionsetup{font=small}
		\includegraphics[width = 0.97\columnwidth]{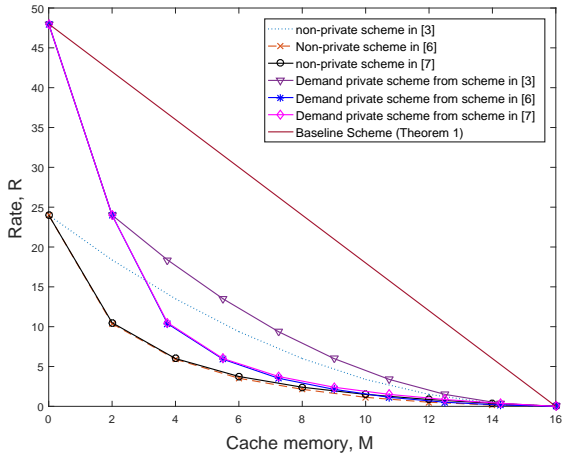}
		\caption{Non-private multi-access coded caching schemes in \cite{ReK,ReK2,CLWZC} and the corresponding demand-private schemes constructed using Theorem \ref{Lifting} and Lemma \ref{sizeprivateset} ($K=24,N=48$ and $L=3$).}
		\label{plot}
	\end{center}
\end{figure}
\begin{exmp}
	Consider the multi-access coded caching scheme with  $K=7$ and $L=5$. For user $U_k,k\in [7]$, $\mathcal{L}^p_k = \left\{\mathcal{Z}_k,\mathcal{Z}_{<k+2>_7},\mathcal{Z}_{<k+4>_7}\right\}$. Consider user $U_1$ with private set $\mathcal{L}^p_1=\{\mathcal{Z}_1,\mathcal{Z}_3,\mathcal{Z}_5\}$. The users $U_2$ and $U_3$ do not have access to $\mathcal{Z}_1$, and the users $U_4$ and $U_5$ do not have access to $\mathcal{Z}_3$. Similarly, the users $U_6$ and $U_7$ do not have access to cache $\mathcal{Z}_5$. In this case, we can manually verify that, $t^* =3$.
\end{exmp}
	\begin{corollary}
		\label{cor1}
		If a memory-rate pair $(M,R)$ is achievable for a $(K,L,N)$ non-private multi-access coded caching scheme satisfying the condition C1, then the memory rate pair $(\tilde{M},\tilde{R}):= \left(M+\ceil{\frac{K-1}{K-L}}(1-\frac{LM}{N}),R\right)$ is achievable for $(K,L,N)$ demand private multi-access coded caching scheme.
		\begin{comment}
		For a $(K,L,N)$ multi-access coded caching scheme, if a memory-rate pair $(M,R)$ is achievable satisfying the conditions C1 and C2, then the memory rate pair $(\tilde{M},\tilde{R}):= \left(M+\ceil{\frac{K-1}{K-L}}(1-\frac{LM}{N}),R\right)$ is achievable with demand privacy.
		\end{comment}
	\end{corollary}
\begin{rem}
	For $L\leq \ceil{\frac{K}{2}}$, the achievability result in Corollary \ref{cor1} coincides with the achievability result in \cite{LWCC}. 
\end{rem}
      %Corollary \ref{cor1} follows directly from Theorem \ref{Lifting} and Lemma \ref{sizeprivateset}.
      
      The rate-memory trade-off of the non-private schemes in \cite{ReK,ReK2,CLWZC} and the corresponding demand-private schemes for $(K=24,L=3,N=48)$ multi-access network is given in Figure \ref{plot}. For $M\geq \ceil{\frac{K-1}{K-L}}$, the additional cache memory used to ensure demand privacy is upper bounded by $\ceil{\frac{K-1}{K-L}}$, i.e., compared to a non-private multi-access scheme achieving the rate $R$, the corresponding demand private scheme needs an additional cache memory, not more than $\ceil{\frac{K-1}{K-L}}$, to achieve the same rate $R$.   
\section{Conclusion}
\label{Conclusion}
	In this work, we addressed the problem of demand-privacy in the multi-access coded caching system. We put forward the idea of the private set of the users in a multi-access network and presented an algorithm to find the private sets of the users. The demand privacy of the users is ensured by placing the keys in those private sets. The main discussion in this paper was restricted to the uncoded placement of the file contents. Investigating the possibility of coded placement is an interesting problem to work on in the future.   
	
	\section{Acknowledgment}
	\label{Ack}
	This work was supported partly by the Science and Engineering Research Board (SERB) of Department of Science and Technology (DST), Government of India, through J.C. Bose National Fellowship to B. Sundar Rajan.

\end{document}